\documentstyle[11pt,twoside,colloqOHP2010,epsfig]{article}
\begin{document}

\title{Implications of the TTV-Detection of Close-In Terrestrial Planets Around
M Stars for Their Origin and Dynamical Evolution}

\author{N. Haghighipour$^1$ and S. Rastegar$^2$}
\affil{$^1$ Institute for Astronomy and NASA Astrobiology Institute, 
University of Hawaii, Honolulu, HI 96825, USA [nader@ifa.hawaii.edu] \\
$^2$ Center for Interdisciplinary Exploration and Research in Astrophysics, 
Northwestern University, 2145 Sheridan Rd., Evanston, IL 60208, USA} 
\begin{abstract}
It has been shown that an Earth-size planet or a super-Earth, in resonance 
with a transiting Jupiter-like body around an M star, can create detectable 
TTV signals (Kirste \& Haghighipour, 2011). Given the low masses of M stars and
their circumstellar disks, it is expected that the transiting giant planet to have
formed at large distances and migrated to its close-in orbit. That implies,
the terrestrial planet has to form during the migration of the giant planet, 
be captured in resonances, and migrate with the giant body to short-period orbits.
To determine the possibility of this scenario, we have studied the dynamics 
of a disk of protoplanetary embryos and the formation of terrestrial planets 
during the migration of a Jupiter-like planet around an M star.
Results suggest that unless the terrestrial planet was also formed at large
distances and carried to its close-in resonant orbit by the giant planet, it is
unlikely for this object to form in small orbits. 
We present the details of our simulations and discuss the implication of the 
results for the origin of the terrestrial planet.
\end{abstract}

\section{Introduction}

In searching for potentially habitable planets, M stars present the most 
promising targets. Because of their small masses, these stars have the greatest reflex 
acceleration due to an orbiting planet.
The low surface temperatures of these stars place their 
(liquid water) habitable zones at distances of approximately 0.1 to 0.2 AU 
(corresponding to orbital periods of $\sim 20$ to 50 days) where the 
precision radial velocity surveys are normally at their optimal sensitivity. 
Given that within the Sun's immediate neighborhood, more 70\% of stars are of 
spectral type M, it is not surprising that for more than a decade, these stars 
have been the subject of research by many authors (Joshi et al. 1997; 
Segura et al. 2005; Boss 2006; 
Scalo et al. 2007; Grenfell et al. 2007; Tarter et al. 2007).

In the past few years, such research has resulted in  the detection of 25
extrasolar planets around 17 M stars. Slightly more than half of these 
planets are Neptune-mass or smaller, consistent with the fact that M stars 
have smaller circumstellar disks and their planets are less massive compared 
to those of G stars. Among these planets are the first Neptune-mass object 
around the star GJ 436 (Butler et al. 2004), the first Earth-size planet 
around the star GJ 876 (Rivera et al. 2005), and the recently discovered 
Earth-like planet in the habitable zone of the star GL 581 (Vogt et al. 2010). 
 
Although majority of currently known planets around M stars have been
detected using the radial velocity technique, these stars have also been 
targets of transit photometry searches. The MEarth project, a robotically
controlled set of eight 40 cm telescopes at Whipple observatory on Mt. Hopkins
in Arizona, is a transit photometry survey that is dedicated to detecting M stars.
This program has been successful in discovering a 6.6 Earth-mass planet around
M star GJ 1214 (Charbonneau et al. 2009).

The transit timing variation method has also been considered as a 
mechanism for detecting small planets around M stars. As shown by 
Kirste \& Haghighipour (2009, 2011), the variations in the transit timing
of a transiting giant planet due to the perturbation of an
Earth-size body or a super-Earth can be large enough to
match the temporal sensitivity of {\it Kepler} space telescope. Figures 1 and 2 show 
samples of the results by these authors.
As shown in figure 1, an Earth-size planet in a 10-day orbit around a 0.32 
solar-mass star produces strong TTVs on a transiting Jupiter-mass planet
when the two objects are in (1:2), (2:3), (5:2), and (2:1) mean-motion
resonances. Figure 2 shows the mean-motion resonances for which an Earth-like
planet in the habitable zone of an M star will produce TTVs of the order
of 10 s or larger on a transiting Jupiter-like body. 

Although the calculations by Kirste \& Haghighipour (2009, 2011) point to the 
detectability of terrestrial planets in systems studied by these authors, 
the low masses of circumstellar disks around M stars cast doubt in the existence 
of their assumed planetary configurations. Computational simulations have indicated
that circumstellar disk around M stars are not massive enough to accommodate
the formation of giant planets, even in orbits as large as that of Jupiter around 
the Sun (Laughlin et al. 2004). The fact that observational surveys have been able to
detected many Jovian-type planets around M dwarfs [e.g. GJ 876 with two Jupiter-like 
planets and a Uranus-mass body in approximately 30, 60, and 120 days orbits 
(Rivera et al. 2005, 2010), or HIP 57050 with a Saturn-mass planet in a 42 days orbit 
(Haghighipour et al. 2010)] suggests that these giant planets were probably formed 
at larger distances, where the disk contained more material, and migrated to their 
current short-period 
orbits. It would therefore be necessary to study how such a migration affects
the formation of terrestrial planets around M stars and their final orbital configuration 
as the giant planet approaches short-period orbits.  

\vskip -3pt
\begin{figure}[ht]
\begin{center}
\epsfig{width=12cm,file=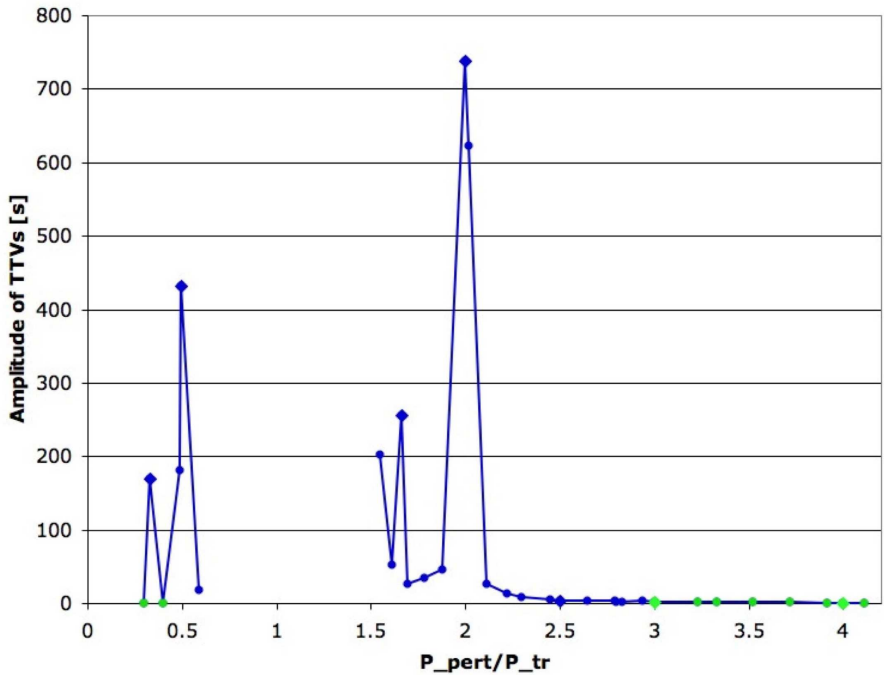}
\caption{Transit timing variations of a 1 Jupiter-mass planet
around a 0.32 solar-mass M star. The perturber is an Earth-sized planet in a 
10-day orbit. The graph shows the values of TTVs for different ratios of the
orbital periods of the two planets. As shown here, when the two planets
are in (1:2), (2:1), (5:3), and (2:3) resonances, the TTVs have values larger 
than 100 sec. Figure from Kirste \& Haghighipour (2011).}
\end{center}
\end{figure}

\begin{figure}[ht]
\begin{center}
\epsfig{width=12cm,file=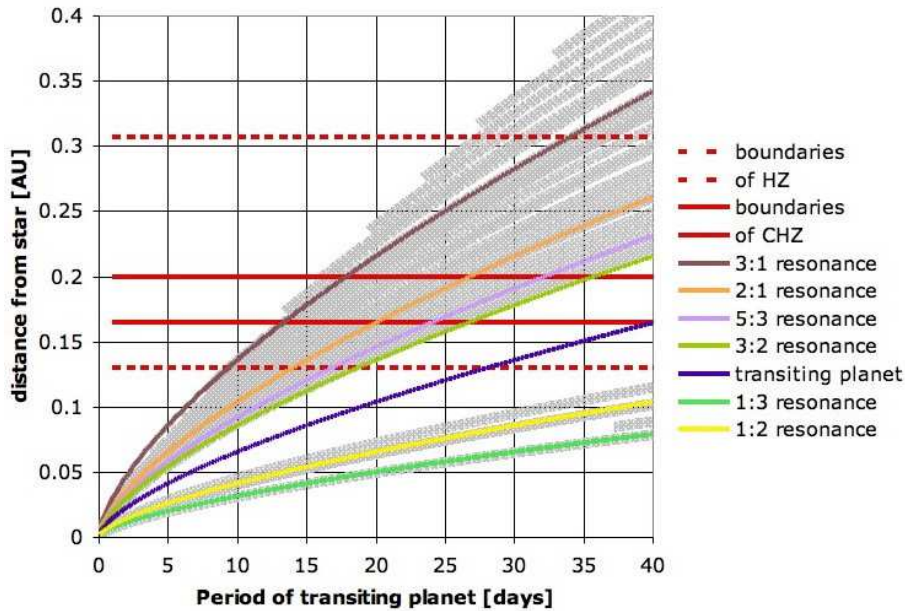}
\vskip -10pt
\caption{Graph of transit timing variation for different resonances between
an Earth-size perturber and a Jupiter-mass transiting planet. The central star
is a 0.36 Solar-mass M star. The habitable zone (HZ) and continuous habitable
zone (CHZ) of the star are shown.The shaded area corresponds to TTVs larger
than 10 sec. A shown here, (1:3), (1:2), and (3:5) resonances produce
large TTVs when the perturber is in the continuous habitable zone.}
\end{center}
\end{figure}

\section{Giant Planet Migration and the Accretion of Embryos}

Terrestrial planet formation in the presence of a migrating giant planet
has been studied by many authors. Examples can be found in the works of
Zhou et al. (2005), Fog \& Nelson (2005, 2007a\&b, 2009), Raymond et al. (2006), 
Mandell et al. (2007), and Kennedy \& Kenyon (2008). As shown by these authors,
a migrating giant planet may capture protoplanetary objects in mean-motion 
resonances and increase their orbital eccentricities to high values.
The latter prevents the accretion of these bodies to larger sizes
by either scattering them to outer distances, or increasing their impact
velocities to values beyond their fragmentation limits. 

If during
the migration of the giant planet, substantial amount of gas still exists,
the combination of gas drag and dynamical friction may prevent the eccentricities
of planetary embryos to reach high values and may facilitate their growth to
larger objects. Simulations by these authors have shown that, 
an Earth-size planet can form 
around a Sun-like star while a Jupiter-mass body migrates through the 
disk of planetary embryos and the system is subject to gas drag. 
At times, the final terrestrial planet was even  
captured in a mean-motion resonance with the giant body and migrated to
close-in orbits (Zhou et al. 2005).

In order to assess the possibility of the existence of planetary systems
studied by Kirste \& Haghighipour (2009, 2011), similar simulations
have to be carried out for a migrating giant planet and a protoplanetary 
disk around an M stars. We note that in the above-mentioned simulations, 
the migration of the giant planet was stopped before it reached 
to very short-period (e.g., 3-day or 4-day) orbits.
Such a termination of the migration was necessary to ensure that 
the terrestrial planets would in fact form, and would not be 
scattered out or crash into the central star. In the system studied by Kirste \&
Haghighipour (2009, 2011), the giant planet revolves around the central star 
in 3-5 day orbits. That suggests, in order to examine the viability of the scenario
presented by these authors, simulations of terrestrial planet 
formation have to be carried out for a migrating giant planet in a disk
of planetary embryos while allowing the giant planet to migrate to  very close-in
orbits.

\section{Numerical Simulations and the Results}

To simulated the formation of terrestrial planets during the migration of 
a Jupiter-like body around an M star, we considered a
model consisting of a star, a protoplanetary disk, and a Jupiter-size planet. 
We assumed the central star to be similar to GJ 876 and have
a mass of 0.32 solar-masses. The protoplanetary disk was considered to be of two types.
Once, similar to Zhou et al. (2005), we randomly distributed 30 planetary embryos,
with masses ranging from 0.1 to 0.5 Earth-masses, in a region between 
0.05 AU (Terquem \& Papaloizou 2007) and 0.5 AU. 
The mutual separations of embryos were chosen to be no smaller than 10 Hill's radii. 
Their eccentricities and inclinations were set to 0.001 and 0.001 deg, respectively.
In the second model, we changed the number of
protoplanets to 40 and distributed them randomly between 0.1 AU  and  0.8 AU. 
The outer edge of the disk in this model was chosen to be equal to 2.7 times the mass 
of the central star as suggested by Kennedy \& Kenyon (2008). In both disk models,
the disk surface density followed an $r^{-1.5}$ profile.

We integrated the motions of planetary embryos
using the N-body integrator MERCURY (Chambers 1999). We modified MERCURY to include
planet migration (Lee \& Peale 2002), gas drag (Raymond et al. 2006), tidal force
(Mardeling \& Lin 2002, 2004), eccentricity damping (Lee \& Peale 2002), and
general relativity (Saha \& Tremaine 1992). Integration were carried out for
both disk
models and for different values of the rate of the migration of the giant planet. 
We assumed that the giant planet was initially at 1 AU and radially migrated with rates
of ${10^{-5}},\, {10^{-6}},\,$ and $10^{-7}$ AU/years. Figure 3 shows the results of 
one of such simulations. The protoplanetary disk in this simulation is of the second
type and the rate of migration is $10^{-7}$ AU/year. As shown here, during
the migration of the giant planet, planetary embryos collid and form bigger objects. 
Collisions were considered to be perfectly inelastic and result in the prefect accretion 
of both bodies. The interaction between embryos and the giant planet caused many of these 
objects to be captured in resonance. However, the latter increased the orbital eccentricities of 
these bodies which eventually resulted in their scattering to large distance. We stopped
the simulation when the giant planet reached the 3-day orbit. All our simulations
showed that no terrestrial planet survived when the giant planet reached short-period orbits.

\section{Conclusions}
Simulations results indicate that although for both disks models and all migration rates,
terrestrial planets were formed in the protoplanetary disk, they did not maintain stability 
and were ejected from the system. The time of the ejection is inversely
proportional to the rate of giant planet migration. Our study suggests that if a terrestrial
planet is detected in resonance with a transiting giant planet around an  M star, 
1) the terrestrial planet is unlikely to have formed in-situ,
2) formation at far distances followed by resonance capture and migration while
in resonance seems to be more viable,
3) the capture probability varies with the migration rate which itself depends on
the mass of the protoplanetary disk.
The latter suggests that slow migration rates and small protoplanetary disks may
in fact facilitate the formation and subsequent resonance capture of a terrestrial
planet with a close-in giant planet around M stars.

\acknowledgments{
NH acknowledges support from NASA Astrobiology Institute (NAI) under cooperative agreement
NNA04CC08A at the Institute for Astronomy, University of Hawaii, NAI central,
and the NASA/EXOB grant NNX09AN05G.}

\eject

\begin{figure}[ht]
\begin{center}
\epsfig{width=9.5cm,file=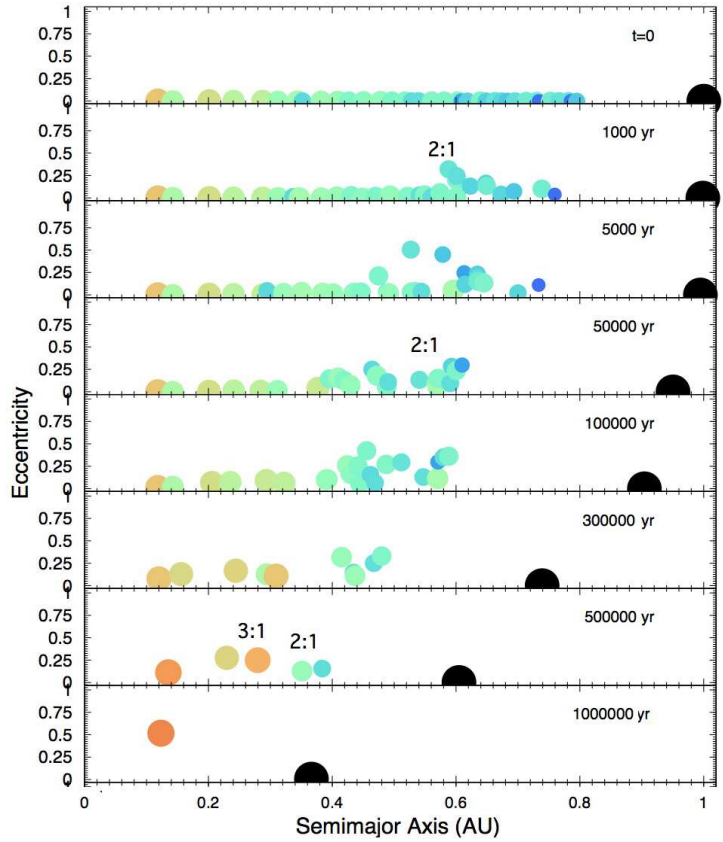}
\vskip -10pt
\caption{Terrestrial planet formation and resonance capture during the migration
of a giant planet (black circle). The disk consists of 40 protoplanets with masses
of 0.1 to 0.5 Earth-masses. The continuation of the simulation shows
that no terrestrial planet survives when the giant planet reaches the 3-day orbit.} 
\end{center}
\end{figure}

\end{document}